\documentclass[a4paper,11pt]{article}
\usepackage{pos}
\usepackage{hyperref}

\title{Two-loop corrections to the Higgs trilinear coupling in models with extended scalar sectors}
 \ShortTitle{Two-loop corrections to the Higgs trilinear coupling}

\author*[a,b]{Johannes Braathen}
\author[a]{Shinya Kanemura}

\affiliation[a]{Department of Physics, Osaka University, Toyonaka, Osaka 560-0043, Japan}

\affiliation[b]{Deutsches Elektronen-Synchrotron DESY, Notkestra\ss{}e 85, D-22607 Hamburg, Germany}

\emailAdd{johannes.braathen@desy.de}
\emailAdd{kanemu@het.phys.sci.osaka-u.ac.jp}

\abstract{The Higgs trilinear coupling provides a unique opportunity to study the structure of the Higgs sector and probe indirect signs of BSM Physics – even if new states are somehow hidden. In models with extended Higgs sectors, large deviations in the Higgs trilinear coupling can appear at one loop because of non-decoupling effects in the radiative corrections involving the additional scalar states. It is then natural to ask how two-loop corrections modify this result, and whether new large corrections can appear again. We present new results on the dominant two-loop corrections to the Higgs trilinear coupling in several models with extended scalar sectors. We illustrate the analytical expressions with numerical examples and show that, while they remain smaller than their one-loop counterparts and do not modify significantly the non-decoupling effects observed at one loop, the two-loop corrections are not entirely negligible – a typical size being 10-20\% of the one-loop corrections.}

\FullConference{%
  40th International Conference on High Energy physics - ICHEP2020\\
  July 28 - August 6, 2020\\
  Prague, Czech Republic (virtual meeting)
}

\begin{document}

\begin{flushright}
OU-HET-1083\\
DESY 20-217
\end{flushright}

\maketitle

\section{Introduction}\vspace{-0.3cm}
In the quest to understand the nature of Beyond-the-Standard-Model (BSM) physics, it is certain that the Higgs trilinear coupling $\lambda_{hhh}$ will play an important role in the near future. A first reason for this is that its determination will allow probing the shape of the Higgs potential away from the electroweak (EW) minimum, which is currently unknown -- at present we only know the location of the EW minimum, given by the Higgs vacuum expectation value (VEV), and the curvature of the potential around it, given by the Higgs mass of 125 GeV. In turn, $\lambda_{hhh}$ also decides the nature and strength of the electroweak phase transition (EWPT). For instance, it is known~\cite{Grojean:2004xa,Kanemura:2004ch} that a deviation of $\lambda_{hhh}$ from its SM prediction of at least 20\% is necessary for the EWPT to be of strong first order, which is itself a requirement for the scenario of electroweak baryogenesis to be successful. 

Additionally, the Higgs trilinear coupling can also serve to probe indirect signs of new physics, even in aligned scenarios~\cite{Gunion:2002zf} where new states are somehow hidden. Indeed, aligned scenarios -- in which the Higgs couplings are SM-like at tree level -- currently seem to be strongly favoured by experimental searches, and non-aligned scenarios could be almost entirely excluded in a foreseeable future, by using the synergy of direct searches at the HL-LHC and indirect searches at lepton colliders -- see for instance Ref.~\cite{Aiko:2020ksl}. One may then ask what is the origin of this alignment. A first option is that alignment can be a consequence of decoupling -- $i.e.$ all BSM states are beyond our experimental reach. Another more interesting option is that alignment can occur even \emph{without decoupling} (possibly because of some symmetry or mechanism). In this latter case, couplings of the Higgs boson -- and in particular $\lambda_{hhh}$ -- can deviate significantly from their SM predictions, because of non-decoupling effects involving BSM states, as found originally in Refs~\cite{Kanemura:2002vm,Kanemura:2004mg}. Moreover, the Higgs trilinear coupling is an ideal candidate to search for such effects because it is currently not well constrained experimentally -- presently the best limits are obtained by ATLAS as $-3.7 < \lambda_{hhh} /\lambda_{hhh}^\text{SM} < 11.5$ (at 95\% confidence level)~\cite{ATLAS:2019pbo} -- but this will be drastically improved at future collider (see $e.g.$ Ref.~\cite{deBlas:2019rxi}). However, one may naturally wonder what happens to the large effects found at one loop once higher-order corrections are included, and whether new huge corrections can appear. 
 
To answer these questions, we computed in Refs.~\cite{Braathen:2019zoh,Braathen:2019pxr} the dominant two-loop corrections to $\lambda_{hhh}$ in several BSM theories with extended scalar sector. Specifically, we considered an aligned scenario of a Two-Higgs-Doublet Model (2HDM), a dark-matter inspired scenario of an Inert Doublet Model (IDM) -- in which the second CP-even mass eigenstates is light and is a dark matter candidate -- and a real singlet extension of the SM that we refer to as ``Higgs-singlet model" (HSM). We refer the reader to section II of Ref.~\cite{Braathen:2019pxr} for details on our conventions for these models. This work was also extended recently in Ref.~\cite{Braathen:2020vwo} for models with classical scale invariance, as well as for massive models with $N$ singlet scalars. In these proceedings, after briefly summarising the setup of our computation, we discuss some examples of numerical results at two loops for the BSM deviation of the Higgs trilinear coupling in the 2HDM and IDM.

\section{Calculational setup}\vspace{-.3cm}

The aim of our calculation is to determine the possible size of two-loop corrections to the Higgs trilinear coupling. For this reason, we choose to employ an effective-potential approximation -- $i.e.$ we derive an effective Higgs trilinear coupling, thereby neglecting subleading effects from external momenta. The steps of our derivation are as follows:\vspace{-.3cm}
\begin{enumerate}
 \item We compute the effective potential $V_\text{eff}$ for the BSM theory at hand. Two-loop contributions are given by one-particule-irreducible vacuum bubble diagrams, and generic $\overline{\text{MS}}$  expressions for these -- applicable to any renormalisable theory -- can be found in Ref.~\cite{Martin:2001vx}.\vspace{-.3cm}
 \item We obtain an effective Higgs trilinear coupling $\lambda_{hhh}$ as the third derivative of the effective potential with respect to the Higgs field $h$, evaluated at the minimum of the potential -- $i.e.$ $\lambda_{hhh}\equiv\frac{\partial^3V_\text{eff}}{\partial h^3}\big|_\text{min.}$. As we calculate corrections to $V_\text{eff}$ in the $\overline{\text{MS}}$ scheme, the results for $\lambda_{hhh}$ are also expressed at first in terms of $\overline{\text{MS}}$-renormalised parameters. \vspace{-.3cm}
 \item Finally, we include the necessary finite counterterms to express our results in terms of physical quantities ($i.e.$ pole masses and physical Higgs VEV), and we also take into account effects from finite wave function renormalisation. This allows us to obtain on-shell scheme results for the Higgs trilinear coupling, which we denote $\hat\lambda_{hhh}$ to distinguish them from the $\overline{\text{MS}}$ results. In this process, we have devised a prescription to renormalise the new mass parameters appearing in BSM models ensuring the proper decoupling of BSM effects~\cite{Braathen:2019zoh,Braathen:2019pxr}. \vspace{-.25cm}
\end{enumerate}
We should also note that we neglect in our work subleading contributions from light scalars (125-GeV Higgs boson and would-be Goldstone bosons), as well as from gauge bosons and light fermions -- in other words, we only consider effects from heavy BSM scalars and the top quark. Lastly, for the 2HDM, we neglect loop-induced deviations from the alignment limit, as these must be small (see $e.g.$ Ref~\cite{Braathen:2017izn}) and as this enables us to evade experimental constraints. 
 
\section{Numerical results}\vspace{-.3cm}
In this section, we present some examples of numerical results for the two-loop corrections to the Higgs trilinear coupling. As our main interest is the possible size of BSM effects, we show our findings in terms of the BSM deviation $\delta R$,  defined as $\delta R\equiv (\hat\lambda_{hhh}^\text{BSM}-\hat\lambda_{hhh}^\text{SM})/\hat\lambda_{hhh}^\text{SM}$. Analytic expressions for $\hat\lambda_{hhh}$ in the SM and the considered BSM theories can be found in Refs.~\cite{Braathen:2019pxr,Braathen:2019zoh}.

\begin{figure}
 \centering
 \includegraphics[width=0.5\textwidth]{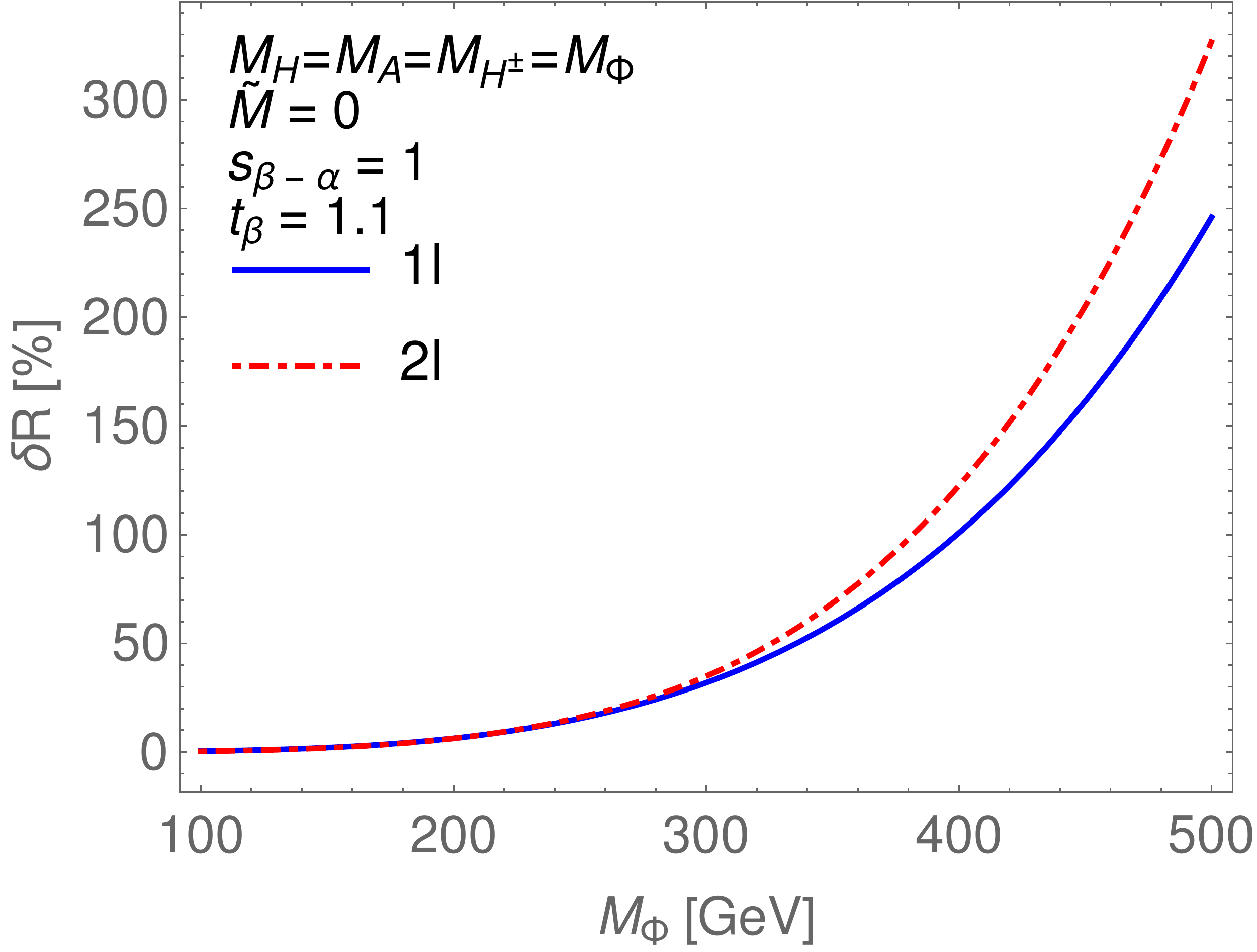}
 \caption{Non-decoupling behaviour of the BSM deviation $\delta R$ in the 2HDM as a function of the degenerate pole mass $M_\Phi$ of the BSM scalars, in the alignment limit $s_{\beta-\alpha}=1$ and for $\tilde M=0$ and $\tan\beta=1.1$. One-loop results are shown in blue, while our new results including two-loop corrections are in red. }
 \label{FIG:2hdm_nondecoup}
\end{figure}

First, we present in figure~\ref{FIG:2hdm_nondecoup} the behaviour of this BSM deviation $\delta R$ as a function of the degenerate mass $M_\Phi$ of the additional scalars of the 2HDM ($H$, $A$, $H^\pm$), taking the alignment limit $s_{\beta-\alpha}=1$, $\tan\beta=1.1$, and setting the BSM mass scale $\tilde M$ to zero (to maximise non-decoupling effects).  One can notice that the two-loop corrections grow faster -- like $M_\Phi^6$ -- than their one-loop counterparts -- that scale as $M_\Phi^4$ -- however the two-loop corrections remain well smaller than the one-loop ones for the entire range of $M_\Phi$ that is considered. For most of this mass range, they amount to 10-20\% of the one-loop corrections, and they reach about 30\% for $M_\Phi\gtrsim 500\text{GeV}$.

\begin{figure}[h]
 \centering
 \includegraphics[width=0.45\textwidth]{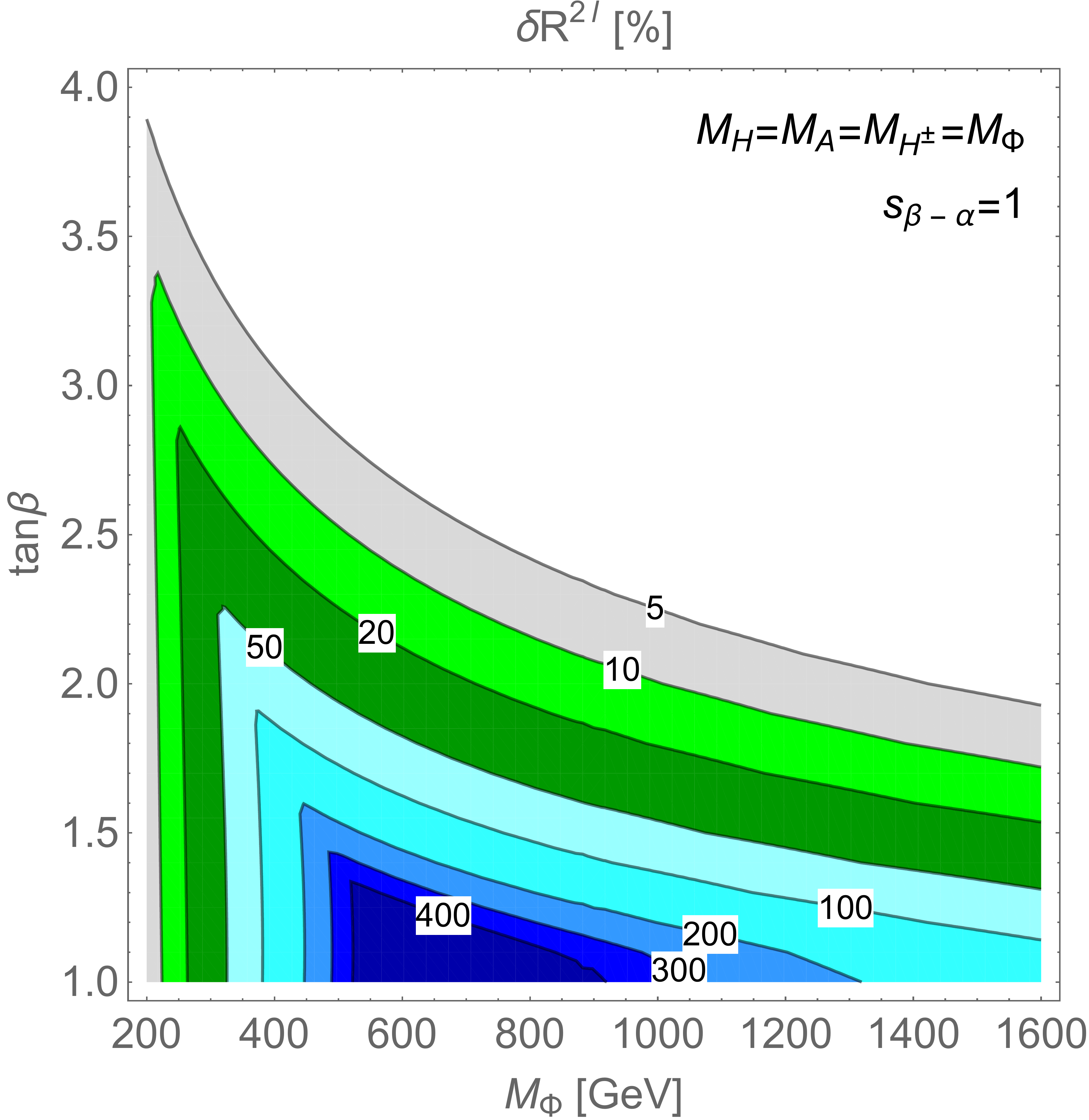}
 \caption{Maximal possible deviation of the Higgs trilinear coupling $\hat\lambda_{hhh}$ computed at two loops in the 2HDM, with respect to the SM prediction, in the plane of $\tan\beta$ and $M_\Phi$, given the requirement of tree-level perturbative unitarity. }
 \label{FIG:2HDM_contourplot} 
\end{figure}

Figure~\ref{FIG:2HDM_contourplot} illustrates the maximal BSM deviation $\delta R$ that can be achieved at two loops while fulfilling the requirement of tree-level perturbative unitarity~\cite{Lee:1977eg,Kanemura:1993hm}, in the plane of $\tan\beta$ and $M_\Phi$. The largest deviations are found for low $\tan\beta$ and intermediate values of $600\text{ GeV}\lesssim M_\Phi\lesssim 800 \text{GeV}$, $i.e.$ when the BSM scalars acquire large masses entirely from the Higgs VEV (in other words $\tilde M=0$). In this region of parameter space, already at one loop, $\hat\lambda_{hhh}^\text{2HDM}$ deviates by as much as 300\% from the SM prediction, and two-loop corrections add an additional effect of order 100\%. However, if one increases $M_\Phi$ further, it becomes impossible to fulfill perturbative unitarity conditions while maintaining $\tilde M=0$, and therefore suppression factors -- $c.f.$ eq. (6.1) in Ref.~\cite{Braathen:2019pxr} -- come into play and reduce the total magnitude of the BSM deviation. Moreover, for larger $\tan\beta$, the constraints from perturbative unitarity become more stringent and hence the allowed BSM deviation diminishes. 
Lastly, we note that the blue-shaded regions (for which $\delta R \gtrsim 50\%$) can be probed at the HL-LHC, while the green-shaded regions ($\delta R\gtrsim 10\%$) will be within reach at future lepton colliders such as the ILC. Detailled studies of the expected accuracy of measurements of the Higgs trilinear coupling at future colliders can be found for instance in Ref.~\cite{deBlas:2019rxi} and references therein. 

\begin{figure}[h]
 \centering
 \includegraphics[width=0.4\textwidth]{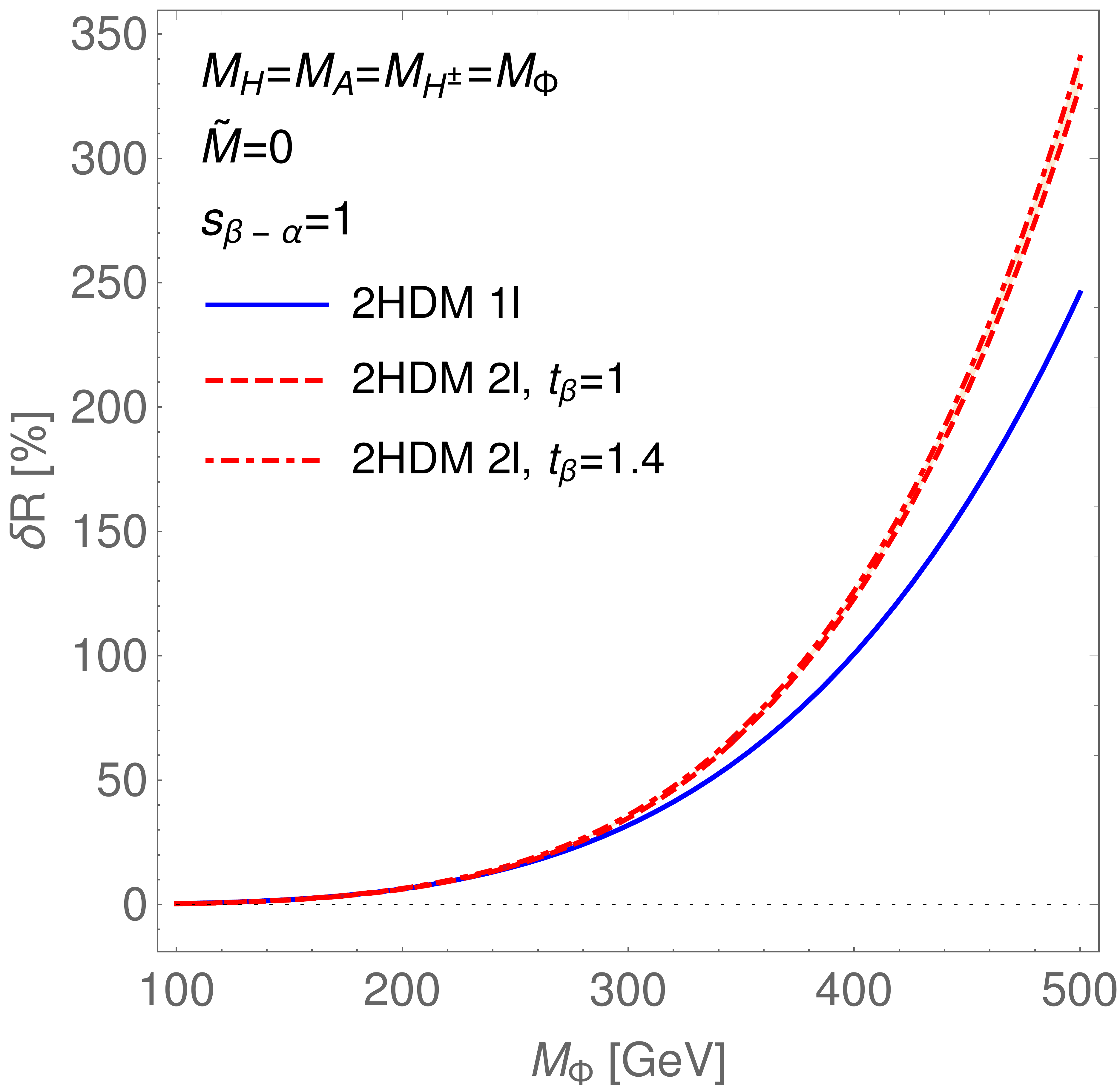}   
 \includegraphics[width=0.4\textwidth]{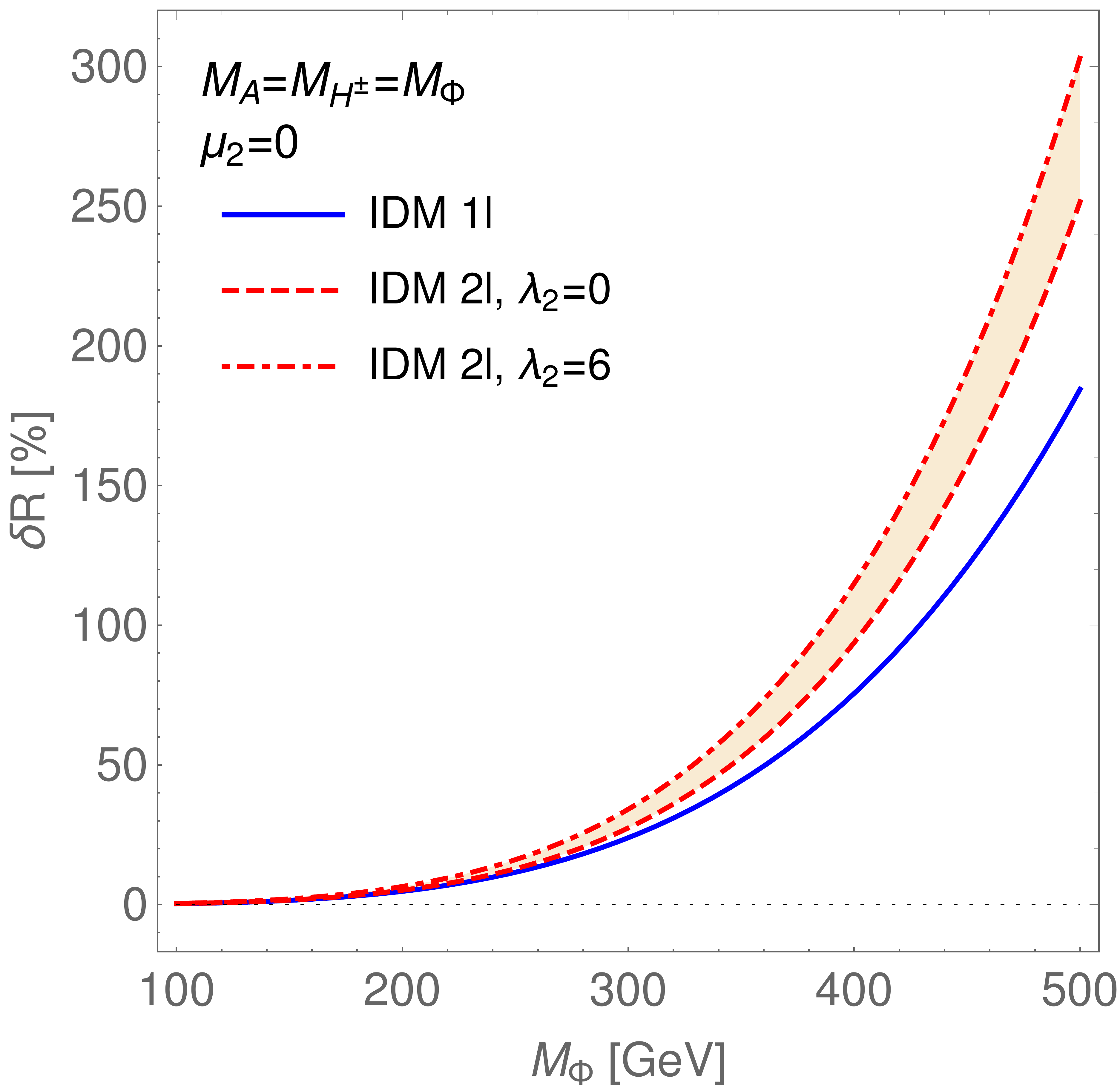}
 \caption{Deviation of the Higgs trilinear coupling from its SM prediction, as a function of the degenerate BSM scalar mass, in the 2HDM (left side) and in the IDM (right side). Blue and red curves correspond respectively to one- and two-loop results. \vspace{-.2cm}}
 \label{FIG:comparison2}
\end{figure}

Finally, in figure~\ref{FIG:comparison2}, we illustrate the possible effect of the additional parameters appearing only two loops in the BSM corrections to $\hat\lambda_{hhh}$ -- respectively $\tan\beta$ for the 2HDM (left) and $\lambda_2$ (the quartic coupling of the second inert doublet) for the IDM (right). For the 2HDM, we vary $\tan\beta$ between 1 and 1.4 -- the latter being the largest value for which perturbative unitarity is maintained for $\tilde M=0$ and $M_\Phi$ up to 500 GeV. Given this very restricted range of allowed values, there are no large effects at two loops from $\tan\beta$. On the other hand, for the IDM, we can vary $\lambda_2$ between 0 (to ensure that the potential is bounded from below) and 6 (to maintain perturbative unitarity up to $M_\Phi=500\text{ GeV}$ for $\tilde\mu_2=0$). This allows large enhancements of $\hat\lambda_{hhh}$ from $\lambda_2$ as shown by the orange shaded region in the right pane of fig.~\ref{FIG:comparison2}. Similar effects are also found with $\lambda_S$ (the singlet quartic coupling) in the HSM~\cite{Braathen:2019pxr}.  

\section{Conclusion}\vspace{-.3cm}
We have presented here results from the two-loop calculations of the Higgs trilinear coupling in a number of BSM theories with extended scalar sectors -- specifically a 2HDM scenario with alignment and the IDM. The two-loop corrections that we obtain amount typically to 10-20\% of one-loop contributions (at most 30\% in the most extreme cases, near the limit where perturbative unitarity is violated). This implies that the non-decoupling effects known to appear at one loop (since Ref.~\cite{Kanemura:2002vm,Kanemura:2004mg}) are not drastically altered. At the same time, our findings also mean that computations of the Higgs trilinear coupling beyond one loop will become necessary for consistent comparison of theoretical and experimental results, given the expected accuracy of its measurement at future colliders (see $e.g.$ Ref.~\cite{deBlas:2019rxi}). Finally, this work illustrates how the precise calculation of Higgs couplings -- $e.g.$ $\hat\lambda_{hhh}$, or its couplings to gauge bosons -- can allow distinguishing aligned scenarios with or without decoupling, by accessing non-decoupling effects.

\acknowledgments \vspace{-.3cm}
This work is, in part, supported by Grant-in-Aid for Scientific Research on Innovative Areas, the Ministry of Education, Culture, Sports, Science and Technology, No. 16H06492 and No. 18H04587. This work is also supported in part by JSPS KAKENHI Grant No. A18F180220. This work is also partly supported by the Deutsche Forschungsgemeinschaft (DFG, German Research Foundation) under Germany’s Excellence Strategy -- EXC 2121 “Quantum Universe” -- 390833306.

\bibliographystyle{JHEP}
\bibliography{BKproc}

\end{document}